
\font\titolino=cmbx10
\font\tsnorm=cmr10
\font\tscors=cmti10
\font\tsnote=cmr9

\font\tscors=cmti10
\font\tscorsp=cmti10
\magnification=1200
\hsize=154truemm
\hoffset=5truemm
\parindent 8truemm
\parskip 3 truemm plus 1truemm minus 1truemm
%
\newcount\notenumber

\def\note{\advance\notenumber by 1 \footnote{$^{\the\notenumber}$}}
\def\beginref{\bigskip
\leftline{\titolino References.}
\nobreak\noindent}
\def\ref#1#2{\noindent\item{\hbox to 25truept{[#1]\hfill}} #2.\smallskip}
\def\beginsection#1#2{%
\bigskip
\leftline{\titolino #1. #2.}
\nobreak\medskip\noindent}
\def\beginack
{\bigskip
\leftline{\titolino Acknowledgments}
\nobreak\medskip\noindent}

\def\sc{\scriptstyle}

\def\Schw{Sch\-warz\-sch\-ild}

\def\da{\dagger}
\def\pf{\pi_f}
\def\pphi{\pi_\varphi}
\def\pf{\pi_f}
\def\dtf{\dot f}
\def\dtvf{\dot\varphi}
\def\dtpf{\dot\pf}
\def\dtpphi{\dot\pphi}
\def\pz{\pi^0}
\def\pu{\pi^1}
\def\d{\partial}

\def\vf{\varphi}

\def\gmn{g_{\mu\nu}}

\def\emn{\eta^{\mu \nu}}

\def\eab{\eta^{\alpha \beta}}
\def\eabb{\eta_{\alpha \beta}}
\def\epsab{\varepsilon^{\alpha\beta}}
\def\real{{\vrule height 1.6ex width 0.05em depth 0ex
    \kern -0.06em {\rm R}}}

\def\sqr#1#2{{\vcenter{\hrule height.#2pt
                       \hbox{\vrule width.#2pt height#1pt\kern#1pt
                       \vrule width.#2pt}
                       \hrule height.#2pt}}}

\def\PRD#1#2#3{{\tscors Phys.\ Rev.} {\bf D#1}, #2 (#3)}
\def\PRL#1#2#3{{\tscors Phys.\ Rev.\ Lett.} \ {\bf #1}, #2 (#3)}
\def\NPB#1#2#3{{\tscors Nucl.\ Phys.} {\bf B#1}, #2 (#3)}

\def\PLB#1#2#3{{\tscors Phys.\ Lett.} {\bf B#1}, #2 (#3)}

\def\IJMPA#1#2#3{{\tscors Int.\ J.\ Mod.\ Phys.} {\bf A#1}, #2 (#3)}
\def\IJMPD#1#2#3{{\tscors Int.\ J.\ Mod.\ Phys.} {\bf D#1}, #2 (#3)}
\def\MPLA#1#2#3{{\tscors Mod.\ Phys.\ Lett.} {\bf A#1}, #2 (#3)}

\def\ANP#1#2#3{{\tscors Ann.\ Physics (N.Y.)} {\bf #1}, #2 (#3)}

%
\def\Hen{M.\ Henneaux, \PRD{54}{959}{1985}}
\def\Nav{J.\ Navarro-Salas, M.\ Navarro and V.\ Aldaya, \PLB
{292}{19}{1992}}
\def\Mic{A.\ Mikovic, \PLB {291}{19}{1992}}
\def\KT{H.A.\ Kastrup and T.\ Thiemann, \NPB{425}{665}{1994}}
\def\Kuchb{K.V.\ Kucha\v r, J.D.\ Romano and M.\ Varadarajan,
\PRD{55}{795}{1997} and references therein}
\def\BH{M.\ Cavagli\`a, V.\ de Alfaro and A.T.\ Filippov,
\IJMPD{4}{661}{1995} and \IJMPD{5}{227}{1996}}
\def\Land{L.D.\ Landau and E.M.\ Lifshitz, {\it The Classical Theory of
Fields} (Pergamon Press, 1962)}
\def\Jauch{See for instance J.M.\ Jauch and F.\ Rohrlich, {\it The Theory of
Photons and Electrons} (Addison-Wesley, 1955) p.\ 103;
C.\ Itzykson and J-B.\ Zuber, {\it Quantum Field Theory} (McGraw-Hill,
1985) p.\ 127}
\def\LGK{D.\ Louis-Martinez, J.\ Gegenberg and G.\ Kunstatter, \PLB
{321}{193}{1994}}
\def\BJL{E.\ Benedict, R.\ Jackiw and H.-J.\ Lee, \PRD{54}{6213}{1996}}
\def\CJZ{D.\ Cangemi, R.\ Jackiw and B.\ Zwiebach, \ANP{245}{408}{1995}}
\def\CJA{D.\ Cangemi and R.\ Jackiw, \PRL {69}{233}{1992}}
\def\CJB{\PRD{50}{3913}{1994}}
\def\CJC{\PLB{337}{271}{1994}}
\def\AER{D.\ Amati, S.\ Elitzur and E.\ Rabinovici, \NPB {418}{45}
{1994}}
\def\Kuma{W.\ Kummer, H.\ Liebl and D.V.\ Vasilevich, ``{\it Exact
Path Integral Quantization of Generic 2-D Dilaton Gravity}'', Report-No: 
TUW-96-28, e-Print Archive: gr-qc/9612012}
\def\Kumb{W.\ Kummer and S.R.\ Lau, ``{\it Boundary Conditions and Quasilocal
Energy in the Canonical Formulation of All 1+1 Models of Gravity}'',
Report-No: TUW-96-27, e-Print Archive: gr-qc/9612021}
\def\Fila{A.T.\ Filippov, in: {\it Problems in Theoretical Physics},
Dubna, JINR, June 1996, p.\ 113}
\def\Filb{A.T.\ Filippov, \MPLA {11}{1691}{1996}; \IJMPA{12}{13}{1997}}
\def\Holl{H.\ Hollmann, \PLB{388}{702}{1996}; 
{\it ``A Harmonic Space Approach to Spherically Symmetric Quantum Gravity''},
Report-No: MPI-PTH-96-53, Oct. 1996, e-Print Archive: gr-qc/9610042}
\def\CGHS{C.\ Callan, S.\ Giddings, J.\ Harvey and A.\ Strominger, \PRD
{45}{1005}{1992}}
%
\null
\rightline{DFTT 20/97}
\rightline{e-Print Archive: hep-th/9704164}
\rightline{February 12, 1997}
\vskip 6truemm
\centerline{\titolino THE BIRKHOFF THEOREM IN THE QUANTUM THEORY}
\vskip 4truemm
\centerline{\titolino OF TWO-DIMENSIONAL DILATON GRAVITY}
\vskip 8truemm
\centerline{\tsnorm Marco Cavagli\`a$^{(a,d)}$,
Vittorio de Alfaro$^{(b,d)}$ and Alexandre T. Filippov$^{(c)}$}
\bigskip
\centerline{$^{(a)}$\tscorsp Dept.\ of Physics and Astronomy, Tufts
University,}
\smallskip
\centerline{\tscorsp Medford, MA 02155, USA.}
\bigskip
\centerline{$^{(b)}$\tscorsp Dipartimento di Fisica
Teorica dell'Universit\`a di Torino,}
\smallskip
\centerline{\tscorsp Via Giuria 1, I-10125 Torino, Italy.}
\bigskip
\centerline{$^{(c)}$\tscorsp Joint Institute for Nuclear Research}
\smallskip
\centerline{\tscorsp R-141980 Dubna, Moscow Region, Russia.}
\bigskip
\centerline{$^{(d)}$\tscorsp INFN, Sezione di Torino, Italy.}
\vskip 5truemm
\vskip 5truemm
\centerline{\tsnorm ABSTRACT}
\begingroup\tsnorm\noindent
In classical two-dimensional pure dilaton gravity, and in particular in
spherically symmetric pure gravity in $d$ dimensions, the generalized
Birkhoff theorem states that, for a suitable choice of coordinates, the
metric coefficients are only functions of a single coordinate.  It is
interesting to see how this result is recovered in quantum theory by the
explicit construction of the Hilbert space. We examine the 
CGHS model, enforce the set of auxiliary conditions that
select physical states \`a la Gupta-Bleuler, and prove that the matrix
elements of the metric and of the dilaton field obey the classical
requirement. We introduce the mass operator and show that its eigenvalue
is the only gauge invariant label of states. Thus the Hilbert space is
equivalent to that obtained by quantum mechanical treatment of the static
case. This is the quantum form of the Birkhoff theorem for this model. 
%
\vfill
\noindent
{\tsnorm PACS: 04.60.-m, 04.60.Kz, 03.70.+k.\hfill}
\break\noindent
{\tsnorm Keywords:} {\it Quantum Gravity, Two-Dimensional models,
Canonical Quantization, Field Theory}
\smallskip
\hrule
\noindent
\leftline{\hbox to 48truemm{Corresponding author:\hfill}Prof.\ Vittorio de 
Alfaro}
\leftline{\hbox to 48truemm{\hfill}Dipartimento di fisica teorica 
dell'Universit\`a}
\leftline{\hbox to 48truemm{\hfill}Via Giuria 1,~~ I-10125 Torino, ITALY}
\leftline{\hbox to 48truemm{\hfill}Tel.\ + 39 - 11 - 670 72 15,~~Fax.\ + 39 
- 11 - 670 72 14}
\leftline{\hbox to 48truemm{\hfill}E-mail: vda@to.infn.it}
\endgroup
\vfill
\eject
\footline{\hfill\folio\hfill}
\pageno=1
\noindent
\beginsection{1}{Introduction}
The quantization of two-dimensional dilaton theories has received much
attention, because of its connection to string theory and also, for some
choices of the potential, to spherically symmetric pure gravitational
$d$-dimensional configurations [1-11]. 

Classically, the two-dimensional pure dilaton theories obey the
generalized Bir\-khoff theorem, namely their solutions depend, in suitable
coordinates, on a single variable; we shall refer to this property as
``staticity''. This is a well known property (see e.g.\ [12,13]). The
famous case is gravity in four dimensions, where, by a suitable choice of
coordinates, a purely gravitational spherically symmetric configuration
can be cast in the form
$$ds^2=-A(r)dt^2+N(r)dr^2+r^2d\Omega^2\,,\eqno(1)$$
where $d\Omega^2$ is the metric of the unit two-sphere and $(r,t)$ are
coordinates defined on $\real^+$ and $\real$ respectively. The
coefficients of the metric are only functions of the radial coordinate $r$:
this is the content of the classical Birkhoff theorem. Now it is
interesting to see how this property appears in the quantum case, where it
reduces the two-dimensional field theory to quantum mechanics. 

In the quantum framework there are two possible approaches. The
two-dimensio\-nal theory can be quantized as a field theory, and then
reduction enforced, or one can quantize the system obtained after the
classical static property has been introduced (direct static
quantization), thus imposing staticity from the beginning [14,15]. In the
latter case one is confronted with a constrained quantum mechanical system
that for the Schwarzschild case has been explicitly solved, and its
Hilbert space determined [14]. 

It is then interesting to study in detail the reduction of the field
theory to quantum mechanics and its connection to the direct static
quantization. In this paper we discuss the CGHS model [4]. We linearize
and implement the constraints \`a la Gupta-Bleuler. This scheme of
quantization allows the algebraic construction of physical states.  In
this scheme, the result is equivalent to the quantization of pure
scalar-longitudinal electrodynamics (apart from the presence of the mass
operator). In this way the staticity property appears explicitly. The
expectation values of the metric and dilaton fields give back the
corresponding classical static formulae. The choice of the quantum state
corresponds to the classical choice of coordinates. 

The only relevant physical fact is the existence of the mass operator,
which is the zero mode of the dilaton field and commutes with the scalar
and longitudinal modes of the D'Alembert fields. The vacuum is thus
characterized by the mass quantum number. The field theory is reduced
to quantum mechanics.

We may conjecture that the same line should be followed in the more
complicated case of general spherically symmetric pure gravity in $d$
dimensions, and in particular in the Schwarzschild case, leading to the
quantum mechanics formulated in [14]. 
\beginsection{2}{Action and Hamiltonian Formalism}
Our starting point is the two-dimensional action
$$S=\int d^2x\,\sqrt{-g}\,\left[\vf R - {\lambda \over 2}
\right]\,,\eqno(2)$$
where $\gmn$ is a two-dimensional metric and $\vf$ is the ``dilaton
field'' (for $R$ we follow the conventions of [16]). 
This model is related by a Weyl transformation to the CGHS [4] model. As 
in [9] we write the two-dimensional metric  as
$$\gmn=\rho\left(\matrix{\alpha^2-\beta^2&\beta\cr
\beta&-1\cr}\right)\,.\eqno(3)$$
Here $\alpha(x_0,x_1)$ and $\beta(x_0,x_1)$ play the role of the lapse
function and of the shift vector respectively; $\rho(x_0,x_1)$
represents the dynamical gravitational degree of freedom.

It is convenient to introduce the variable $f=\ln \rho.$ Using the ansatz
(3) the action (2) can be written in the Hamiltonian form (see [9]) as
$$S=\int d^2x\,\left[\dot f\pi_{f}+\dtvf\pphi-\alpha
{\cal H}-\beta{\cal P}\right]\,,\eqno(4)$$
where $\pi_f$ and $\pphi$ are the conjugate momenta to $f$ and $\vf$
respectively, and $\cal H$ and $\cal P$ are the super-Hamiltonian and the
super-momentum: 
$$\eqalignno{&{\cal H}=\pf\pphi+f'\vf'-2\vf''+{\lambda\over 2} e^f\,,
&(5a)\cr
&{\cal P}=2\pf'-\pphi\vf'-\pf f'\,.&(5b)\cr}$$
The Hamiltonian equations of motion are
$$\dtvf=\alpha\pf-\beta\vf'\,,~~~~
\dtf=\alpha\pphi-\beta f'-2\beta'\,,\eqno(6a)$$
$$\dtpphi={\d ~\over\d x_1}(\alpha f'+2\alpha'-\beta\pphi)\,,~~~~
\dtpf=-{\lambda\over 2} \alpha e^f +{\d ~\over \d x_1}
(\alpha\vf'-\beta\pf)\,,\eqno(6b)$$
and obviously the constraints
$${\cal H}=0, ~~~~~~~~~~~~{\cal P}=0\,.\eqno(7)$$
We may also define a functional $M$ [7, 12, 13] of $f$ and $\vf$ conserved
under time and space translations (analogous to the \Schw\ mass).  In our
notations $M$ is given by
$$M~=~{\lambda\over 2} \vf ~+~e^{-f}(\pf^2-\vf'^2)\,.\eqno(8)$$
It is straightforward to prove that $\dot M=M'=0$ using the
equations of motion and the constraints.

The Birkhoff classical staticity can be stated as follows. We may
set $\alpha=1$ and $\beta= 0$ and introduce the coordinates
$$u={1\over 2}(x_0+x_1)\,,~~~v={1\over 2}(x_0-x_1)\,;\eqno(9)$$
the two-dimensional line element corresponding to the metric
tensor (3) becomes
$$ds^2=4\rho(u,v)dudv\,.\eqno(10)$$
A metric of this form is static if and only if [12] $\rho$ can be cast in
the form
$$\rho(u,v)=h(\Psi){da(u)\over du}\, {db(v)\over dv}, ~~~~~
\Psi\equiv a(u)+b(v)\,,\eqno(11)$$
where $a$ and $b$ are two suitable functions. This will be useful later.

In the variables (9) the equations of motion and constraints read [12]
$$\eqalignno{&\d_u\d_v f=0\,,~~~~
\d_u\d_v\vf+ {\lambda\over 2} e^f =0\,,&(12a)\cr
&\d_u\left(e^{-f}\d_u\vf \right)=\d_v\left(e^{-f}\d_v\vf
\right)=0\,,&(12b)}$$
where (12b) coincides with ${\cal H}\pm{\cal P}=0$. The solution of these 
equations is static:
$$\eqalignno{&\rho\equiv e^f={dF\over d\Psi}\d_u\Psi\d_v\Psi\,,&(13a)\cr
&\lambda\vf=F(\Psi)=C_0 e^{-\lambda \Psi/2}\,
+\, 2M \,,&(13b)\cr
&\d_u \d_v \Psi\,=\,0\,.&(13c)}$$
One sees that $M$ appears as a zero mode of the field $\vf$. The freedom
of choosing $\Psi$ as a solution of (13c) is related to the
reparametrizations of Eq.\ (11). 

Now let us recall the  canonical free field formalism that will be the
starting point for the quantum theory. 
\beginsection{3}{Canonical Transformation to free fields}
Let us use the transformation [9]
$$\eqalign{&A_0={2\over\lambda}e^{-f/2}\left(\pf\cosh{\Sigma}-
\vf'\sinh{\Sigma}\right)\,,~~
\pz=-\lambda e^{f/2}\cosh{\Sigma}-\lambda A_1'\,,\cr
&A_1={2\over\lambda}e^{-f/2}\left(\pf\sinh{\Sigma}-
\vf'\cosh{\Sigma}\right)\,,~~
\pu=\lambda e^{f/2}\sinh{\Sigma}+\lambda A_0'\,,\cr}\eqno(14)$$
where
$$\Sigma={1\over 2}\int^x dx'\pphi(x')\,.\eqno(15)$$
The above transformation is
canonical. Using (14) the two constraints become
$$\eqalignno{&{\cal H}={1\over
2\lambda}\pi^\alpha\pi_\alpha+{\lambda\over 2}A'^\alpha A'_\alpha=0\,,
&(16a)\cr
&{\cal P}=-\pi^\alpha A'_\alpha=0\,.&(16b)\cr}$$
We introduce the metric and the Levi Civita tensors as
$$\eta^{\alpha \beta}=\left(\matrix{1&0\cr 0&-1\cr}\right)\,;~~~~~
\epsab=\left(\matrix{0&1\cr -1&0\cr}\right)\,.\eqno(17)$$
The functional $M$ defined in (8) is represented as
$$M=M_0+{\lambda\over 4}\int_{b}^x dx' ~\bigl(\epsab\pi_\alpha A_\beta~
+~\lambda A^\alpha A'_\alpha\bigr)\,.\eqno(18)$$
Here $M_0$ must be a constant since $M$ is independent of $x_0$; it
corresponds to a constant in $\vf$ (zero mode). On the equations of motion
we have $M=M_0$. $\dot M$ and $M'$ are proportional to ${\cal
C^{\alpha}}$ defined in next (19). 

The constraints (16) can be cast in a simpler form. Indeed,
Eqs.\ (16) are equivalent to the linearized constraints
$${\cal C}^\alpha\equiv\pi^\alpha-\lambda\epsab A'_\beta=0\,.\eqno(19)$$
The sign ambiguity arising in the process of linearization is removed
using the canonical transformation (14). Finally the general system
described by the action (2) is equivalent to the system described by the
Hamiltonian density
$$H\,=\,{\cal H}\,+l_\alpha{\cal C}^\alpha\,,\eqno(20)$$
where $l_\alpha$ are two Lagrange multipliers that imply ${\cal
C}^\alpha=0$ and thus \ ${\cal H}=0$ and ${\cal P}=0$.

Let us discuss how staticity can be recovered from (20). The equations
of motion are 
$$\lambda \dot A_\alpha=\eabb\pi^\beta\,,~~~~
\dot\pi^\alpha=\lambda\eab A_\beta''\,\eqno(21)$$
and the solution is
$$ A_\alpha=U_\alpha(u)+V_\alpha(v)\,.\eqno(22)$$
Now let us implement the constraints (19), that correspond to
$$\eab \d_{\alpha} A_{\beta}=0,~~\epsab \d_{\alpha} A_{\beta}=0,
~~{\rm or,}~~
U_0(u)=U_1(u),~~V_0(v)=-V_1(v).\eqno(23)$$
It is easy to prove that the above solution is static and coincides with
(13). To this aim we need to obtain $f$ and $\vf$ as functions of
$A_{\alpha}$ inverting (14): 
$$\rho\equiv e^f={2\over\lambda}{\cal H}- 2
A'^\alpha A'_\alpha+{2\over\lambda}\epsab\pi_\alpha A_\beta'\,,~~~~~~~
\vf={2M\over\lambda}-{\lambda\over 2}
A^\alpha A_\alpha\,.\eqno(24)$$
Substituting Eqs.\ (22,23) in (24), and using ${\cal H}=0$, one obtains
$$\eqalignno{&\rho\equiv e^f\,=\,4\lambda
{dU_0(u)\over du}\, {dV_0(v)\over dv}\,,
&(25a)\cr
&\vf\,=\,{2M \over \lambda}- 2\lambda U_0(u)V_0(v)\,.&(25b)\cr}$$
The above solution coincides with (13), with
$$C_0 e^{- \lambda \Psi/2}=-2\lambda^2 U_0(u) V_0(v)\,.\eqno(26)$$
The property of staticity of the classical solution is thus represented by
the conditions (23), to be implemented in the canonical quantization. 
\beginsection{4}{Quantization}
The quantization starts from the introduction of the 
Lagrangian\note{\tsnote From now on we set $\sc\lambda=1$. There is no
real loss of generality while the formulae become more elegant.}
$${\cal L}= {1 \over 2} \d_{\mu}A_{\alpha} \d_{\nu}A_{\beta}
\emn \eab\,.\eqno(27)$$
The commutation relations are
$$\bigl[ A_{\alpha}(x),A_{\beta}(y) \bigr]= -\eabb
\int {d^2k \over 2\pi} \delta(k^2) \varepsilon(k_0)
e^{ik(x-y)}\,. \eqno(28)$$
The field expansion is
$$A_{\alpha}= \int_{-\infty}^\infty {dk\over 2\sqrt{\pi \omega}}
\bigl\{b_{\alpha}(k)e^{-i\omega x_0+ikx_1}
+b^\dagger_{\alpha}(k)e^{i\omega x_0- ikx_1}\bigr\}\,,\eqno(29)$$
where $\omega=|k|$. From (22) we obtain
$$\eqalign{&U_{\alpha}= \int_0^\infty {dk \over 2\sqrt{\pi k}}
\bigl\{ a_{\alpha}(k)e^{-2iku} +a^\dagger_{\alpha}(k)e^{2iku}\bigr\}\,,\cr
&V_{\alpha}=\int_0^\infty {dk \over 2\sqrt{\pi k}}
\bigl\{b_{\alpha}(k)e^{-2ikv}
+b^\dagger_{\alpha}(k)e^{2ikv}\bigr\}\,,\cr}\eqno(30)$$
and $a_\alpha(k)=b_\alpha(-k)$, $k>0$. Consequently ($k>0$) the
non-vanishing commutators are
$$\bigl[a_{\alpha}(k),a^\dagger_{\beta}(k')\bigr] = \eabb
\delta(k-k')\,,~~~~
\bigl[b_{\alpha}(k),b^\dagger_{\beta}(k')\bigr] = \eabb
\delta(k-k')\,.\eqno(31)$$
It follows that
$$\bigl[U_{\alpha}(u_1),U_{\beta}(u_2)\bigr] =
-{i\over 4} \eabb \varepsilon(u_1-u_2),~~~
\bigl[V_{\alpha}(v_1),V_{\beta}(v_2)\bigr] =
-{i\over 4} \eabb \varepsilon(v_1-v_2).\eqno(32)$$
The operator $M_0$ that is the zero mode of $\vf$ commutes with all the
annihilation and creation operators since by Eqs.\ (14) it is not
contained in the fields $A_{\alpha}$. This will be very important later.

The classical constraints (19,23) cannot be implemented operatorially,
since as operator equations they are in contrast with the quantization
rules (28-32). A way of solving for the Hilbert space is to introduce
an indefinite metric in the space of states as in the Gupta-Bleuler
procedure [17]. Following the usual lore we proceed defining the vacuum
as
$$a_{\alpha}(k)|0>=0\,,~~~~~~~~b_{\alpha}(k)|0>=0\,. \eqno(33)$$
This leads to negative norm states. Now we implement the constraints by
requiring that, for each oscillation mode, physical states be selected
by
$$\bigl( a_0(k)-a_1(k) \bigr)\, |\Psi>\,=\,0\,,~~~~
\bigl( b_0(k)+b_1(k) \bigr) \, |\Psi>\,=\,0\,.\eqno(34)$$
Let us introduce
$$q_a= a_0-a_1\,, ~~~~~~~~q_b= b_0+b_1\,,~~~~~~~~
\bigl[q_{a,b}(k),q_{a,b}^\dagger(k')\bigr]=0\,,\eqno(35)$$
and the states $|\{n_a,n_b\}>$ defined as
$$|\{n_a,n_b\}>\equiv q^{\da}_a(k_1)...q^\da_a(k_{n_a})\,
q^{\da}_b(k'_1)...q^\da_b(k'_{n_b})|0>\,.\eqno(36)$$
These states have zero norm if $n_a\,\not=0$ or $n_b\,\not=0$.  The 
general solution of the constraints is then
$$|\Psi>\,=\,\sum_{n_a,\,n_b}\int d^{n_a}k \int d^{n_b}k'~
C_{n_a n_b}(k_1,...k_{n_a};k'_1,...k'_{n_b}) |\{n_a,n_b\}>\,.
\eqno(37)$$
The norm of this state is
$$<\Psi|\Psi>\,=\,|C_{00}|^2\,. \eqno(38)$$
One can check that the classical constraints hold for expectation
values.  After some algebra one finds
$$<\Psi|:{\cal H}:|\Psi>=0\,,~~~~
<\Psi|:{\cal P}:|\Psi>=0\,,\eqno(39)$$
where the normal ordering with annihilation operators on the right must
be used.  It is easy to check that all matrix elements of $\cal
H$ and $\cal P$ among physical states vanish.

Analogously one can calculate the matrix elements of $\rho$, of the mass
$M$, and of the scalar field $\vf$. Using (24) and (37) the expectation
value of $\rho$ is
$$<\Psi|:\rho(u,v):|\Psi>= 4 {dF(u)\over du}\, {dG(v)\over
dv}\,,\eqno(40)$$
where
$$\eqalign{&F(u)=\int {dk\over 2\sqrt{\pi k}}\,\left(
C_{00}{}^* C_{10}(k) e^{-2iku}~+~
C_{00} C_{10}(k){}^* e^{2iku}\right)\,,\cr
&G(v)=\int {dk\over 2\sqrt{\pi k}}\,
\left(C_{00}{}^* C_{01}(k) e^{-2ikv}~+~
C_{00} C_{01}(k){}^* e^{2ikv}\right)\,.\cr}\eqno(41)$$
The result (40) is analogous to the classical relation (25a); we have 
of course
$$F(u)G(v)=<\Psi|U_0(u)V_0(v)|\Psi>\,.\eqno(42)$$
Note that $<\Psi|:\rho(u,v):|\Psi>$ has the form
$$<\Psi|:\rho(u,v):|\Psi>=
h\bigl(a(u)+b(v)\bigr)\,{da(u)\over du}\,{db(v)\over dv}\,,\eqno(43)$$
which is the essence of classical staticity.

Let us now consider the operator $M$, Eq. (18). The quantity
$I$ that is the integrand in (18)
classically vanishes. In the quantum case, each term in $I$ contains one
of the operators $q_{a,b}$ or $q^{\dagger}_{a,b}$. Adopting a normal
ordering, the matrix elements of $I$ between physical states vanish. This
corresponds to the classical property. So,
$$<\Psi_2|M|\Psi_1>=<\Psi_2|M_0|\Psi_1>\,.\eqno(44)$$
Further, the operator $M_0$ commutes with the creation and annihilation
operators of $A_\alpha$, since $M_0$ is the zero mode of the field
$\vf$. So we must characterize the vacuum by a further quantum number: 
$$M_0|0;m>=m|0;m>. \eqno(45)$$
Eq.\ (45) is of the utmost interest. There are infinite vacua, differing
by the eigenvalue of $M_0$. The only gauge invariant label of a state is
$m$. This result is similar to the case of the \Schw\ metric discussed in
[14], where staticity was imposed from the beginning, reducing the problem
to quantum mechanics, and states were labeled by the eigenvalues of the 
mass operator.

Finally, the expectation value of $\vf$ reads
$$\eqalign{<\Psi;m|:\vf(u,v):|\Psi;m>\,&=\,2m-2
<\Psi;m|U_0(u) V_0(v)|\Psi;m>\cr
&=\,2m-2F(u)G(v)\,.}\eqno(46)$$
in analogy to (25b).

We conclude with two interesting remarks. The first is that the roles of
$A_0$ and $A_1$ can be interchanged, i.e.\ the sign in Eq.\ (28) can be
changed,  because the condition (16a) shows that the choice of the
``right'' metric field is irrelevant. The operators $q_a$ and $q_b$ will
again contain one operator with wrong metric and one with right metric;
nothing changes in the construction (37) of the physical states.

Second remark: our quantization rule (28) amounts to assuming that $x_0$
is time, namely that the canonical equal $x_0$ commutators for $A_0$ hold.
The determination of the physical states is actually independent of which
coordinate is chosen as time. Indeed, let us suppose $x_1$ to be the
timelike variable, and proceed by canonical $x_1$ quantization for $A_0$.
In that case the rule (28) is suitably modified. In the rules (31) the
commutators of the $b_{\alpha}$ change sign: now $b_0$ has wrong metric
while $b_1$ has the correct one. Again, in $q_b$ there appears one
operator with the right and one with the wrong metric and the construction
of physical states, Eq.\ (36), remains unchanged. 
\beginsection{5}{Conclusions}
Classically, taking into account the constraints (23), all the field
theory tells us is just that there is a single free field whose degrees
of freedom correspond to reparametrization of the static coordinate,
Indeed, a choice for $U_0(u),\, V_0(v)$ defines $\Psi$, and the different
choices correspond to different solutions of (13c). In the quantum
theory, the physics contained in $A_\alpha$ is pure gauge, equivalent to
free electrodynamics of longitudinal and scalar photons, and in this
respect the state $|\Psi>$ conveys the information correspondent to the
classical case, see (40) and (46). 

What is physically important is the eigenvalue of the constant operator
$M_0$, Eq.\ (45). The vacuum has a quantum number: the eigenvalue of the
mass operator. Thus the theory is reduced essentially to quantum
mechanics, while the rest is coordinate reparametrization. One may
conjecture that this mechanism is at the basis of the dimensional
reduction for all the quantum field models for which classically the
Birkhoff theorem holds.  When a general potential appears in (3) the
problem is the existence and identification of the canonical
transformation, analogous to (14), that leads to free fields. If it is
so, the quantum mechanics [14] that is obtained quantizing the
spherically symmetric metric (1)  in its static form contains
the same physical information as the quantum field theory. 
\beginack
One of us (M.C.) acknowledges a foreign grant by the University of
Torino and support by the Angelo Della Riccia Foundation, Florence,
Italy.
\beginref
\item{[1]} \Hen.
\item{[2]} \Nav.
\item{[3]} \Mic.
\item{[4]} \CGHS.
\item{[5]} \CJA; ~\CJB; ~\CJC.
\item{[6]} \AER.
\item{[7]} \LGK.
\item{[8]} \KT.
\item{[9]} \BJL; ~\CJZ.
\item{[10]} \Kuchb.
\item{[11]} \Kuma.
\item{[12]} \Fila; ~\Filb.
\item{[13]} \Kumb.
\item{[14]} \BH.
\item{[15]} \Holl.
\item{[16]} \Land.
\item{[17]} \Jauch.
\vfill \eject \bye